\documentclass{aa}

\usepackage{graphics}

\begin{document}

   \thesaurus{03     
              (13.25.2;  
               11.19.1;  
               11.09.1)}  
   \title{BeppoSAX observation of NGC 7674: a new reflection-dominated Seyfert 2 galaxy}

   \author{G. Malaguti\inst{1}, G.G.C. Palumbo\inst{1,2}, M. Cappi\inst{1,3},
           A. Comastri\inst{4}, C. Otani\inst{3}, M. Matsuoka\inst{3}, M. Guainazzi\inst{5},
           L. Bassani\inst{1}, \and F. Frontera\inst{1,6}
          }

   \offprints{G. Malaguti (malaguti@tesre.bo.cnr.it)}

   \institute{ITESRE/CNR, via Piero Gobetti 101, I-40129 Bologna, Italy \and 
              Dipartimento di Astronomia, Universit\`a di Bologna, via Zamboni 33, I-40126 Bologna, Italy \and
              The Institute of Physical and Chemical research (RIKEN), 2-1, Hirosawa, Wako, Saitama 351-01, Japan \and
              Osservatorio Astronomico di Bologna, via Zamboni 33, I-40126 Bologna, Italy \and
              ASI SAX/SDC c/o Nuova Telespazio, via Corcolle 19, I-00131, Roma, Italy \and 
              Dipartimento di Fisica, Universit\`a di Ferrara, via Paradiso 12, I-44100 Ferrara, Italy}
   \date{Received / accepted}

   \titlerunning{BeppoSAX observation of NGC 7674: a new reflection-dominated Seyfert 2 galaxy}
   \authorrunning{G. Malaguti et al.}

   \maketitle

   \begin{abstract}

    The Seyfert 2 galaxy NGC 7674 has been observed within the BeppoSAX Core Programme 
    with the onboard 
    narrow field instruments between 0.1 and 100 keV. The broad-band spectrum shows 
    four most relevant spectral components: a) a soft excess below $\sim$2 keV; b) a prominent 
    (EW$\sim$1 keV) Fe line; c) a flat ($\Gamma_{\rm obs}\sim1.1$) 2--10 keV continuum; 
    d) a 4.5$\sigma$ detection above 13 keV. The flat power law spectrum can be very well
    explained within the current AGN unified models assuming a steep ($\Gamma\sim2$) intrinsic
    spectrum scattered by warm gas for the low energy band and totally reflected by 
    optically thick
    cold matter (plausibly a molecular torus) for the high energy band. The case of NGC 7674
    adds to the increasing number of so called ``Compton-thick'' Seyfert 2 galaxies 
    in which the direct emission is totally absorbed and the X-ray luminosity is thus 
    at least one or two orders of magnitude larger than what inferred from the observed flux.

   \end{abstract}

%

\section{Introduction}

Unification models of Seyfert galaxies, at least in their simplest formulation,
postulate the presence of a geometrically and optically thick gas and dust torus
which, at a distance of several to several tens of parsecs from the nucleus
(Krolik, Madau, and Zycki 1994), hides the primary source and the broad line forming region (BLR).
The orientation of the torus is independent of the host galaxy, and the observed differences 
between Seyfert 1 and Seyfert 2 nuclei are thus to be ascribed simply to the angle formed by 
the line of sight direction and the axis of the torus.
After the seminal work by Antonucci \& Miller (1985) various pieces of evidence
have been accumulating in favour of the unification model of Seyfert galaxies 
(see Antonucci 1993 for a review of the subject): 
detection of broadened emission lines in 
optical spectropolarimetry observations of several Seyfert 2 galaxies (Antonucci \& Miller 1985;
Miller \& Goodrich 1990; Tran 1995; Young et al. 1996) interpreted as scattering
of the BLR emission by warm material placed above the torus; 
biconical structures in the light of the forbidden {\sc [oiii]} line at 5007 ${\rm \AA}$ 
(e.g. Tadhunter and Tsvetanov 1989) which is expected if the
ionizing radiation field from the nucleus is anisotropic; large ($10^{23}$ cm$^{-2}$ or greater)
absorbing column densities observed in the hard X-ray spectra of Seyfert 2 galaxies
(Awaki 1997).
 
In the optical and UV bands the nuclear emission is visible only if 
the line of sight does not intercept the torus. In the X-ray band, the interaction 
between electromagnetic radiation and matter is dominated by photoelectric absorption up
to $\simeq10$ keV and by Compton scattering at higher 
energies. When the line of sight intercepts the torus (Seyfert 2) 
primary X-rays are able to leak through it (Compton thin source) 
if the column density N$_{\rm H}$ is less than $\sim1.5\times10^{24}$ cm$^{-2}$.
When N$_{\rm H} >\;{\rm few}\;\times10^{24}$ cm$^{-2}$ the primary X-rays
cannot escape and are either promptly absorbed via photoelectric
interaction, or first Compton downscattered and eventually photoabsorbed
making the nucleus invisible (Compton thick source) in direct emission.

Emission from Compton thick Seyfert 2 galaxies can be detected by means of the reflection
caused by the visible inner surface of the torus and/or via scattering by the material responsible
for producing the broad lines observed in polarized light.
The detection of these sources in X--rays is hampered because of their
faintness ($\la {\rm few} \times 10^{-12}$ erg cm$^{-2}$ s$^{-1}$) and only a few of them
are currently known (Matt 1997 and references therein). Their expected observable features in X-rays
 are: a 2--10 keV continuum flatter than the canonical 
slope observed for Seyfert 1, a strong (equivalent width (EW) $\sim$ 1 keV or higher when
measured against the reflected continuum) Fe K$_\alpha$ feature around 6.4 keV, 
and a reflection ``hump'' at energies $>10$ keV (e.g. Ghisellini, Haardt, and Matt 1994). 

NGC 7674 is a $z=0.029$ Seyfert 2 nucleus hosted in a face on 
spiral galaxy (Sab) with asymmetrical 
arms and a tidal connection to a nearby compact elliptical galaxy. 
Broad H$_\alpha$ and H$_\beta$ 
components in polarized flux were first observed by Miller \& Goodrich (1990),
and then by Young et al. (1996). The spectropolarimetric observations suggest
that NGC 7674 is the only known Seyfert 2 galaxy with hidden broad line region 
for which dust 
scattering is the dominant cause of the observed nuclear polarization (Tran 1995).

Preliminary analysis of the GINGA observation of NGC 7674 (Awaki et al. 1991) gave a flat 
($\Gamma\sim1.5$) 2--10 keV
spectrum with no indication for a Fe line (EW $<$ 110 eV). 
However, subsequent analysis indicated that the observed count rate was 
below the 3$\sigma$ rms fluctuation of the Cosmic X-Ray Background between
2 and 10 keV in the eight detectors of the Large Area Counter 
onboard GINGA (Smith and Done 1996).
As a consequence, the 2--10 keV flux of $4\times10^{-12}$ erg cm$^{-2}$ s$^{-1}$ 
previously reported by Awaki et al. (1991) has to be considered only as a 2$\sigma$ 
upper limit (Smith, private communication).
In the following, detailed measurement of the X-ray spectral characteristics of NGC 7674
obtained by BeppoSAX is reported. 

\section{Observation and data reduction}

The BeppoSAX X-ray observatory (Boella et al. 1997a), a major programme
of the Italian Space Agency with participation of the Netherlands Agency for 
Aereospace Programs, was launched on April 30th 1996 from Cape Canaveral
in Florida. The payload instruments are characterized by a wide spectral coverage 
and consist of
four co-aligned Narrow Field Instruments (NFI) plus two Wide Field Cameras 
perpendicular to the
axis of the NFI and looking in opposite directions. 

The NFI include 
a Low Energy Concentrator Spectrometer
(LECS; Parmar et al. 1997), three Medium Energy Concentrator Spectrometers 
(MECS; Boella et al. 1997b), 
a High Pressure Gas 
Scintillation Proportional Counter (HPGSPC; Manzo et al. 1997), and a Phoswich 
Detector System (PDS; Frontera et al. 1997).
LECS and MECS are instruments with imaging capabilities (angular 
resolution $\sim$ 1$\arcmin$) and operate in the 0.1--10 keV and 1.5--10 keV spectral 
bands respectively. In the overlapping energy interval the effective area 
of the LECS is about one third of the three MECS (which is around 150 cm$^2$ at 6 keV).
HPGSPC and PDS are instruments functioning via rocking collimators and cover the 4--120
keV and 15-300 keV energy bands respectively. In the overlapping energy band PDS is the most
sensitive (by a factor $\sim4-8$) while HPGSPC is optimized for spectral resolution.
%

   \begin{figure}
      \resizebox{\hsize}{!}{\includegraphics{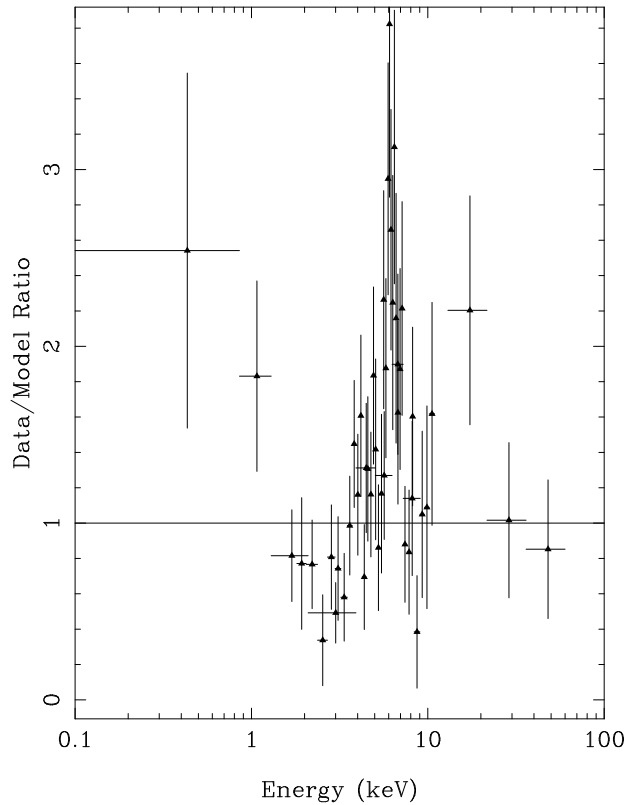}}
      \caption[]{Data to model ratio in the 0.1-100 keV energy
band for a single power law model.}
         \label{res}
   \end{figure}
BeppoSAX NFI pointed at NGC 7674 from November 25th to November 26th 1996 for about 
$1.1\times10^5$ s of effective observing time. The on-source time for the LECS was $\sim$40 ksec
since this instrument is switched on only during satellite night time,
while the on-source time for PDS and HPGSPC was $\simeq$50\% of the effective time due to
rocking collimator functioning.

LECS and MECS good time intervals were selected imposing the minimum 
elevation angle above Earth's 
limb to be $>5^\circ$ and the
instrument to be in its nominal configuration\footnote{Information 
regarding background subtraction 
and BeppoSAX NFI data reduction can be obtained from  
{\em http://www.sdc.asi.it/software/}}.
Given the weak 2--10 keV source flux (see next section) LECS and MECS spectra 
were extracted from a 2$\arcmin$ radius region around the centroid
of the source and the appropriate response matrices were generated.
The same region was used to extract background spectra from blank sky observation
files. The use of local instead of blank sky backgrounds did not make significant 
difference. 

The PDS consists of four phoswich units. They operate in collimator mode,
with two of them pointing to the source while the other two point $\pm210'$ 
away. The two pairs switch on and off source every 96 seconds. The net source spectra
have been obtained by subtracting the `off' to the `on' counts.
PDS source data have been multiplied by a factor 1.42 to account for the mismatch
in the absolute normalization with the MECS (Matt et al. 1997, and references therein).
Leaving this number free to vary would only slightly change the best fit
parameters, without affecting the main results of the present analysis.
The net source count rate in the PDS was $0.13\pm0.03$ cts/s between 13 and 60 keV,
which gives a significant detection also after conservative 
subtraction of the systematic
residuals which are currently evaluated $\simeq0.02$ cts/s in the 13--200 keV band 
(Guainazzi and Matteuzzi 1997).

The net source count rate (in units of $10^{-3}$ cts/s) was $2.9\pm0.3$ in the LECS
(0.1--9 keV), $4.7\pm0.3$ in the three MECS (2--10 keV), while it was below detection 
limit in the HPGSPC.
Neither the soft 
($\sim$0.1--2 keV) nor the hard (2--10 keV) energy band data show significant variability. 
Statistical $\chi^2$ tests against 
constancy gave a probability greater than 25\%, thus consistent with a constant source.
The spectra were thus extracted from the overall observing period and were rebinned to 
have a minimum of 20 counts in each channel.

\section{Spectral analysis}

The spectral analysis has been performed by means of the {\sc XSPEC 9.0} package,
and using the instrument response matrices released by the BeppoSAX Science Data 
Centre in January 1997. 
Figure \ref{res} shows the ratio of the broad band spectrum between 0.1 and 100 keV to a 
simple power law ($\Gamma\simeq0.5$) model. Residuals indicate the presence of
a significant soft excess below 2 keV, a strong line emission around 6.4 keV,
and excess emission in the 13--30 keV band. These clear spectral features make 
NGC 7674 an interesting case for investigating and testing reflection models.
All the quoted errors correspond to 
90\% confidence intervals for one interesting parameter ($\Delta\chi^2$ of 2.71).

\subsection{The 2-10 keV spectrum}
The MECS spectral data in the 2--10 keV energy band were first fitted with an 
absorbed power law model, 
plus a Gaussian narrow ($\sigma=0$ fixed) line to take into account
the 6.4 keV feature, with all other parameters left free to vary. 
The best fit (model \#1 in Table 1)
parameters are a photon index $\Gamma\simeq1.1$,
an absorption column density $N_{\rm H}\simeq6\times10^{22}$ cm$^{-2}$, and a 
gaussian line centered 
(in the reference frame of the emitting source) around
6.4 keV with
an equivalent width of $\sim$1 keV. 
The fit improves significantly ($>$98\% via F-test) if the line is allowed to broaden
and gives a width of $\sigma\simeq0.4$ keV and an equivalent width of $\sim$2 
keV (model \#2).
Unless extreme iron abundances are assumed (about an order of magnitude larger than the cosmic
value), the
observed iron emission line intensity 
cannot be explained by transmission through the measured absorption 
column density which would predict an EW$<$100 eV (Makishima 1986, Ptak et al. 1996). 
Such a strong line, together with the unusual flatness of the continuum, is therefore 
a very strong evidence that the observed hard X-ray
spectrum is reflection-dominated.
\begin{table*}
  \centering
  \caption[line]{Upper panel: MECS fit with a power law plus different models;
lower panel: pure reflection fits.}
  \vspace*{.5cm}
  \begin{tabular}{c ccc lll l}
\hline\hline \\ \\
\multicolumn{8}{c}{2--10 keV -- Single absorbed power-law model} \\ \hline \\
Model \# & \multicolumn{2}{c}{$\Gamma$} &  N$_{\rm H}$                     & E(FeK)  & $\sigma$(FeK)& EW(FeK) & $\chi^{2}/\nu$\\ 
         &        &                     & ($\times10^{22}\;{\rm cm}^{-2}$) & (keV)   & (keV)        & (eV)    &               \\ \hline \\
1 & \multicolumn{2}{c}{$1.12^{+0.34}_{-0.32}$}&$6.4^{+7.3}_{-4.5}$   &$6.39^{+0.18}_{-0.13}$      &      0$_{\rm (fixed)}$ &$1030^{+790}_{-431}$ & 32.4/28 \\ \\ \hline \\
2 & \multicolumn{2}{c}{$1.06^{+0.37}_{-0.32}$}&$4.6^{+7.6}_{-3.8}$   &$6.51^{+0.19}_{-0.17}$      & $0.41^{+0.27}_{-0.16}$&$2070^{+1450}_{-930}$& 26.1/27 \\ \\ \hline \\
   \multicolumn{8}{c}{0.1--100 keV --  Soft component + Pure reflection component} \\ \hline \\
Model \# & \multicolumn{2}{c}{$\Gamma_{\rm soft}$} & $\Gamma_{\rm hard}$   & E(FeK) & $\sigma$(FeK)& EW(FeK) & $\chi^{2}/\nu$ \\ 
         & \multicolumn{2}{c}{kT}                                &                       & (keV)  & (keV)        & (eV)    &                \\ \\ \hline \\
3a &  \multicolumn{2}{c}{$2.87^{+0.53}_{-0.49}$}      &$1.92^{+0.21}_{-0.21}$ &$6.45^{+0.18}_{-0.17}$&$0.26^{+0.27}_{-0.22}$&900$^{+470}_{-299}$& 36.5/38  \\ \\ 
3b &  \multicolumn{2}{c}{$0.85^{+0.55}_{-0.27}$} keV& $2.06^{+0.23}_{-0.23}$&$6.45^{+0.23}_{-0.17}$&$0.23^{+0.27}_{-0.23}$&757$^{+423}_{-252}$& 37.3/38  \\ \\ 
\hline \\
4 &\multicolumn{2}{c}{$1.92^{+0.28}_{-0.27}$} &$\Gamma_{\rm hard}\equiv\Gamma_{\rm soft}$ &  $6.48^{+0.21}_{-0.17}$ & $0.31^{+0.26}_{-0.19}$ &$1070^{+740}_{-448}$ & 39.5/39 \\ \\
\hline \\
5 &\multicolumn{2}{c}{$2.91^{+0.60}_{-0.67}$} &$1.97^{+0.27}_{-0.16}$ &  $6.33^{+0.14}_{-0.16}$ & 0$_{\rm (fixed)}$  & $524^{+224}_{-194}$ &35.9/37 \\
  &                       &                                          & & $6.90^{+0.60}_{-0.70}$ & 0$_{\rm (fixed)}$  & $260^{+111}_{-97}$  & \\ \\ \hline \\
6 &\multicolumn{2}{c}{$2.88^{+0.56}_{-0.58}$} &$1.93^{+0.22}_{-0.21}$   & 6.40$_{\rm (fixed)}$ & 0$_{\rm (fixed)}$  &831$^{+263}_{-232}$& 36.4/39  \\ 
  &           &                       &                                 & 7.06$_{\rm (fixed)}$ & 0$_{\rm (fixed)}$  &103$^{+74}_{-38}$  &           \\ \\ 
\hline\hline \\
\end{tabular} 
\end{table*}

\subsection{The soft excess}

The low statistics at energies below $\simeq$2 keV do not allow to
firmly discriminate among different models tested. In fact,
the excess detected in the LECS data at $E<2$ keV is well modeled both by a
steep ($\Gamma\sim2.9$) power law absorbed by the galactic column density 
in the direction of the source 
(N$_{\rm H} = 5.31 \times 10^{20}$ cm$^{-2}$; Dickey and Lockman [1990])
and by a Raymond-Smith model with
solar abundance and kT$\sim$0.9 keV (models \#3a and \#3b).
Moreover, 
an acceptable fit is also obtained assuming an
electron scattering model with $\Gamma_{\rm soft}\equiv\Gamma_{\rm hard}$ (model
\#4). The possible implications of the scattering model
are discussed in section 4.

\subsection{The broad-band reflection-dominated spectrum}
%

   \begin{figure}
      \resizebox{\hsize}{!}{\includegraphics{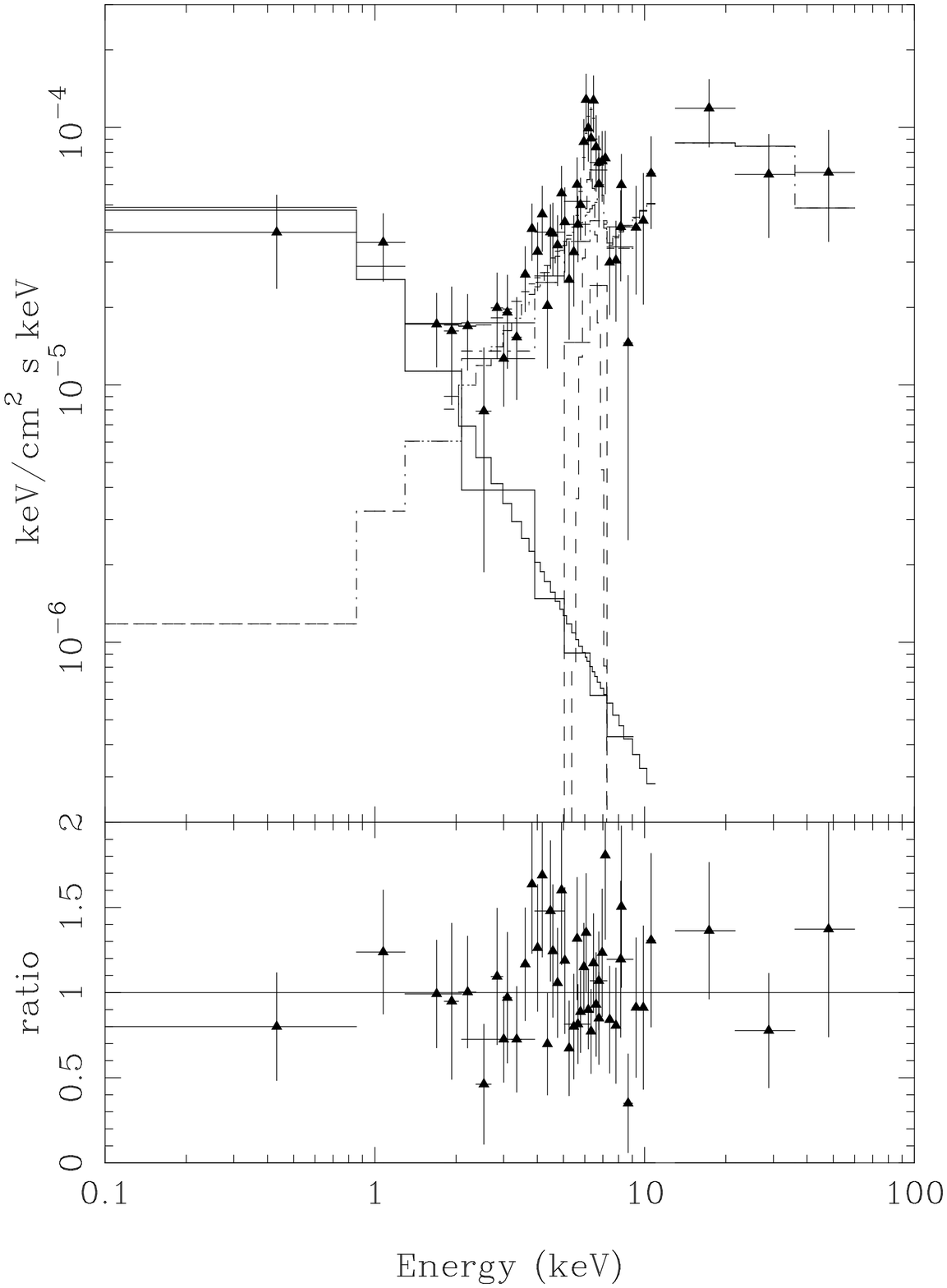}}
      \caption[]{Unfolded best fit model (model \#3a) consisting of a steep power law at low energies
plus a ``pure'' reflection component and associated FeK line at higher energies. }
         \label{bf}
   \end{figure}
The broad band (0.1--100 keV) spectrum has been fitted with a soft component (see section 3.2)
plus a pure reflection continuum 
resulting from a power law illumination of cold and thick material plus a gaussian line.
The use of this model is justified by the fact that the emission line centroid 
energy is consistent with K shell fluorescence
from neutral Iron and has an
EW of the order of what is expected in the case of 
a pure reflection continuum (e.g. Matt, Perola, and Piro 1991; Ghisellini, Haardt, 
and Matt 1994; Krolik, Madau, and Zycki 1994).
In this reflection model ({\sc plrefl} in {\sc xspec})
the only free parameter is the relative normalization of the reflected component to the
direct one. 
If this parameter is left free to vary, only a lower limit ($\simeq40$) 
is obtained, thus confirming that the observed spectrum is dominated by reflection.  
Therefore a purely reflected spectrum (i.e. no direct component) was chosen. 
The best fit gives in this case (model \#3a) an intrinsic photon index
$\Gamma\simeq1.9\pm0.2$ consistent with the average values of Seyfert 1 galaxies
(Nandra and Pounds 1994).
The $2-10$ keV observed flux is $5\times10^{-13}$ erg cm$^{-2}$ s$^{-1}$ corresponding
to an observed luminosity of $2\times10^{42}$ erg s$^{-1}$.
The observed $\sim6.4$ keV emission feature can also be fitted with a more complex model.
An acceptable fit (model \#5) is in fact obtained by adding to the Fe neutral line
a second emission feature.
The best-fit energy centroid of this second line (6.90 keV)
is consistent both with H-like Fe (6.97 keV) or with a blend of He- (6.70 keV) plus H-like Fe.
This indicates that, certainly, reflection from thick cold
matter plays an important role. Nonetheless, contribution from reflection caused by ionized
matter cannot be ruled out on the basis of the present data.
Finally, an acceptable fit (model \#6) is also obtained assuming both K$_\alpha$
and K$_\beta$ emission to be present in the form of narrow lines.
The best-fit equivalent widths are consistent with the K$_\beta$/K$_\alpha$
ratio expected for neutral iron (0.135 [Weast 1987]).

\section{Discussion}

The 2-10 keV flat spectrum together with the high equivalent width of the Fe line
strongly suggest that the observed spectrum of the Seyfert 2 galaxy NGC 7674 is 
reflection dominated. 
The observed luminosity, $L_{\rm Obs}^{\rm 2-10\;keV}=2\times10^{42}$ erg s$^{-1}$ is, 
therefore, expected to be much lower than the intrinsic one. 
Assuming a scattering model (model \#4), the best fit luminosity of the 
soft component extrapolated in the 2--10 keV band is 
$L_{\rm Scatt}^{\rm 2-10\;keV}\simeq4\times10^{41}$ erg s$^{-1}$. 
If the material that electron-scatters the nuclear radiation is confined
in a cone-shaped region above the torus, then
$L_{\rm Scatt}$ is linked to
the nuclear intrinsic luminosity $L_{\rm Int}$ by the relation
$L_{\rm Scatt}=L_{\rm Int}\tau(\Omega/2\pi)$, where 
$\tau$ and $\Omega$ are the mirror optical depth and subtended solid angle respectively.
Assuming NGC 1068 values of $\simeq10^{-3}$ and 0.25 for $\tau$ and $\Omega/2\pi$ 
respectively (Iwasawa, Fabian, and Matt 1997),
then $L_{\rm Int}$ is estimated to be $\sim10^{45}$ erg s$^{-1}$.
It is interesting to note that this value is very close to 
the intrinsic {\sc [oiii]} luminosity of $\sim6\times10^{44}$ erg s$^{-1}$
evaluated from the observed flux corrected for the
reddening in the narrow line region (Maiolino, private communication).
The fact that also the far infrared integrated $8-120$ $\mu$m luminosity  
L$_{\rm FIR}=1.6\times10^{45}$ erg s$^{-1}$
(Spinoglio and Malkan 1989) is of the same order of magnitude
suggests that
both L$_{\rm [OIII]}$ and and L$_{\rm FIR}$ can be   
good indicators of the intrinsic X-ray luminosity.

If the reflection is caused by cold materials at the inner surface of the torus 
(Ghisellini, Haardt, and Matt 1994; Krolik, Madau, and Zycki 1994), then
only a fraction $f_{\rm R}$ of the reflecting surface is visible. 
This fraction can be estimated from the intrinsic luminosity.
Using the notation
of Iwasawa, Fabian, and Matt (1997), the observed luminosity can be written
as $L_{\rm Scatt}=f_{\rm R}\eta L_{\rm Int}$
where $\eta$ is the albedo for a isotropic illumination. 
Using the above values for $L_{\rm Scatt}$
and $L_{\rm Int}$ and assuming $\eta=0.022$, $f_{\rm R}$ becomes equal to $\sim$0.1, which is 
consistent within a factor of 2 of what found for NGC 1068.

Pure reflection spectra have been 
observed in X-rays by the ASCA satellite in at least few other Seyfert 2 
galaxies such as NGC 6552 (Fukazawa et al. 1994; 
Reynolds et al. 1994), NGC 6240 (Kii et al. 1997),
the Circinus galaxy (Matt et al. 1996), and
NGC 1068 which has also been confirmed by BeppoSAX 
(Matt et al. 1997, and references therein).
In NGC 7674, even if some contribution from a warm reflection component 
cannot in principle be excluded, as demonstrated by the acceptable fit obtained with a blend of
ionized lines (model \#5),
the observed X-ray emission is certainly dominated by reflection from cold matter.
In fact, on statistical basis (better $\chi^{2}_{\nu}$, see Table 1) and self-consistency arguments,
the most convincing explanation for the possible higher energy line
is K$_{\beta}$ emission from neutral iron (model \#6).
This is contrary to what observed in NGC 1068 and NGC 6240 where the ionized reflection
component is much stronger, but similar to what observed in the Circinus galaxy and
NGC 6552. 
Such differences may have different physical interpretations.
They could be ascribed to an orientation effect (more face-on than NGC 1068 and NGC 6240) with 
respect to the line of sight of the torus, or to the condition (e.g. low scattering efficiency) 
of the warm mirror itself.
In conclusion, the case of NGC 7674 adds to the increasing number of Compton-thick 
Seyfert 2 galaxies,
thus suggesting that a dedicated study of a complete sample of optically selected 
Seyfert 2 galaxies may 
allow the discovery of a significant number of Compton-thick objects as indicated by recent
BeppoSAX results (Salvati et al. 1997; Maiolino et al. 1997).

\begin{acknowledgements}
We thank Dr. G. Matt for helpful discussions, and Dr. D. Smith who provided us with the Ginga data.
This research has made use of SAXDAS linearized and cleaned event files produced at
the BeppoSAX Science Data Centre. G.M., G.G.C.P., M.C., A.C., and L.B. acknowledge 
financial support from the Italian Space Agency. 
\\
\end{acknowledgements}

\end{document}